\documentstyle[preprint,
               epsfig,amssymb,aps]{revtex}

\begin{document}
 
%\preprint{}\draft
 
\title{Clustering and Correlations at the Neutron Dripline}
 
\author{N.A.~Orr\thanks{e-mail: {\tt orr@caelav.in2p3.fr}},
F.M.~Marqu\'{e}s\thanks{e-mail: {\tt marques@caelav.in2p3.fr}}}

\address{Laboratoire de Physique Corpusculaire,
IN2P3-CNRS, ISMRA et Universit\'e de Caen,\\ Boulevard Mar\'echal Juin, 14050 Caen cedex, France}

%\date{\today}

\maketitle

%%%%%%%%%%%%%%%%%%%%%%%%%%%%%%%%%%%%%%%%%%%%%%%%%%%%%%%%%%%%
%%%  Abstract  %%%
%%%%%%%%%%%%%%%%%%

\begin{abstract}
Some recent experimental studies of 
clustering and correlations 
within very neutron-rich light nuclei are reviewed.  In particular, the 
development of the novel probes of neutron-neutron
interferometry and Dalitz-plot analyses is presented through
the example of the dissociation of the two-neutron halo system $^{14}$Be.
The utility of high-energy proton radiative capture is illustrated using
a study of the $^{6}$He(p,$\gamma$)
reaction.
A new approach to the production and detection of bound neutron clusters
is also described, and the observation of events with the characteristics expected
for tetraneutrons ($^{4}$n) liberated in the breakup of 
$^{14}$Be is discussed.  The prospects for future work, including systems
beyond the neutron dripline, are briefly outlined. 
\end{abstract}

%%%%%%%%%%%%%%%%%%%%%%%%%%%%%%%%%%%%%%%%%%%%%%%%%%%%%%%%%%%%
%%%  Main text (in English)  %%%
%%%%%%%%%%%%%%%%%%%%%%%%%%%%%%%%

\section{Introduction}

Clustering, which has long been known to occur along the line
of beta stability,
also appears in exotic forms as the drip-lines are approached \cite{VonO,Freer}. 
The most spatially extreme form of clustering is that
exhibited by neutron haloes
which appear as the ground states of very weakly bound nuclei at the limits of particle 
stability \cite{Tanihata,Nunes}. Perhaps the most intriguing 
of the halo systems are the Borromean two-neutron halo nuclei ($^{6,8}$He, 
$^{11}$Li, 
$^{14}$Be and $^{17}$B), in which the two-body subsystems -- core-$n$ and $n$-$n$ -- are 
unbound. Such 
behaviour naturally gives rise to the question of the correlations between the 
constituents. Even in the case of the most studied 
of these nuclei, $^{6}$He and $^{11}$Li, little is known in this respect.
Here aspects of an experimental programme which has involved 
the development 
of a number of novel tools specifically aimed at studing clustering 
and correlations in halo and related neutron-rich systems is outlined.

In the first part of this review (section 2) the nature of these correlations
is explored through the application of the techniques of 
interferometry and 
Dalitz-plot analyses to 
kinematically complete measurements of dissociation.
In the case of breakup of a halo system, both the reaction mechanism and
final state interactions (FSI) come into play.  As described below, neutron-neutron
interferometry and Dalitz-plot anlayses exploit the FSI between the fragments in
the exit channel.  In a complementary approach, 
radiative capture -- in which the outgoing photon should not be affected by FSI --
has been investigated 
as a probe of clustering in nuclei far from stability (section 3).
The example described here is that of the $^{6}$He(p,$\gamma$) reaction.

On a more speculative note, the production and detection of bound multineutron 
clusters in the breakup
of very neutron-rich secondary beams is explored in section 4.  
This approach exploits the possibility that
multineutron halo nuclei and other very neutron-rich systems contain components
of the wavefunction in which the neutrons exist in a relatively compact cluster-like
configuration. 
A method is introduced for the direct detection of neutral clusters and 
the results
obtained from an analysis of data acquired with beams of $^{11}$Li and $^{14}$Be
is presented.  

By way of a conclusion, the prospects for future work using the 
techniques presented 
here are outlined in section 5.  The possibilities for extending 
these studies to explore the
structure of systems beyond the neutron dripline are also briefly described.

\section{Correlations in Two-Neutron Halo Nuclei}

As introduced above, we have explored the spatial configuration of halo
neutrons at breakup through the 
application of the technique of intensity interferometry;
an approach first developed in their pioneering work on stellar 
interferometry by 
Hanbury-Brown and Twiss in 
the 1950's and 60's \cite{HBT} and later extended to source size measurements in 
high energy
collisions \cite{Gol60}.  The principle behind the technique is as follows:  
when identical particles are emitted 
in close proximity in space-time, the wave function
of relative motion is modified by the FSI and quantum statistical
symmetries \cite{Boa90} --- in the case of halo neutrons the overwhelming effect is that of 
the FSI \cite{FMM00}. Intensity interferometry relates this
modification to the space-time separation of the particles at emission as a 
function of the four-momenta of the particles through the correlation function 
$C_{\rm{nn}}$, which is defined as,

\begin{equation}
 C_{\rm{nn}}(p_1,p_2)=\frac{d^2n/dp_1dp_2}{(dn/dp_1)\,(dn/dp_2)} \label{e:C12}
\end{equation}

where the numerator is the measured two-particle distribution and the denominator the 
product of the independent single-particle distributions \cite{FMM00}.  
As is generally the case, the single-particle distributions
have been generated via event mixing.  Importantly, in the 
case of halo neutrons
special consideration must be given to the strong residual correlations \cite{FMM00},
whilst experimentally care needs to be taken to eliminate
cross talk \cite{Mar00}.

As a first step, measurements of breakup 
of $^{6}$He, $^{11}$Li and 
$^{14}$Be by a Pb target \cite{E295,Lab01} were analysed \cite{FMM00} (the details of 
the experiment may be found in
refs \cite{E295,Lab01,FMM00}).  The choice of a high-Z target was made to 
privilege
Coulomb induced breakup, whereby the halo neutrons may in a first approximation 
act as spectators and for which simultaneous emission may be 
expected to occur.  The correlation functions derived from the data, assuming 
simultaneous emission, were compared to
an analytical formalism based on a Gaussian source \cite{soviet}.
Neutron-neutron separations of $r_{nn}^{RMS}=5.9\pm1.2$~fm ($^{6}$He), 
$6.6\pm1.5$~fm ($^{11}$Li) and $5.6\pm1.0$~fm ($^{14}$Be) were thus extracted.  These 
results appear to
preclude any strong dineutron component in the halo wavefunctions at breakup; a
result which, for $^{6}$He, is in line with the radiative capture study reported in the
following section.
It is interesting in this context to compare these results to the RMS neutron-proton
separation of 3.8~fm in the deuteron (the only bound two nucleon system).
The same analysis has been applied to dissociation of $^{14}$Be by a C target,
in order to investigate the influence of the reaction mechanism \cite{FMM01}.  A 
result which hints at a somewhat larger separation, $r_{nn}^{rms}=7.6\pm1.7$~fm, was 
obtained. This raises the question as to whether simultaneous emission can be
assumed a priori. In principle, the analysis of the correlation function in two
dimensions, transverse and parallel to the total momentum of the pair, would
allow for the unfolding of the source size and lifetime \cite{soviet}. Such an
analysis requires a much larger data set than presently available. 
The two-neutron halo, however, is far less complex than
the systems usually studied via interferometry (for example, heavy-ion 
collisions \cite{Boa90}).
Moreover, the simple three-body nature of the system breaking up suggests
that any delay in the emission of one of the neutrons will arise
from core-$n$ FSI/resonances in the exit channel, a process that may be expected to be
enhanced for nuclear induced breakup.

Correlations in three-particle decays are commonly encountered in particle
physics and are typically analysed using plots of the
squared invariant masses of particle pairs $(M_{ij}^2,M_{ik}^2)$, with
$M_{ij}^2=(p_i+p_j)^2$; a technique developed by the Australian physicist Richard Dalitz
in the early 1950's \cite{Dal53,Example}. In Dalitz-plot representations, 
FSI or resonances lead to a
non-uniform population of the surface within the kinematic boundary defined by
energy-momentum conservation and the decay energy. 
In the present case, the core-$n$-$n$ system exhibits a distribution of decay
energies ($E_{\rm{d}}$). The $E_{\rm{d}}$ associated 
with each event
will thus lead to a different kinematic boundary, and the resulting plot
containing all events cannot be easily interpreted. A
normalised invariant mass has thus been introduced \cite{FMM01}, 

\begin{equation}
 m_{ij}^2 = \frac{M_{ij}^2-(m_i+m_j)^2}{(m_i+m_j+E_{\rm{d}})^2-(m_i+m_j)^2}
\end{equation}

which ranges between 0 and 1 
(that is, a relative energy $E_{ij}=M_{ij}-m_i-m_j$ 
between 
0 and $E_{\rm{d}}$) 
for all events and exhibits a single kinematic boundary. 
Examples of how $n$-$n$ and core-$n$ FSI may manifest themselves 
in the Dalitz plot for the decay of $^{14}$Be are illustrated in Fig.~2, whereby 
events have been simulated according to the simple interacting phase-space
model described in ref. \cite{FMM01}. The inputs were an 
$E_{\rm{d}}$ distribution following that measured \cite{Lab01}, the 
$C_{\rm{nn}}$ obtained with the C target, and a
core-$n$ resonance with $\Gamma=0.3$~MeV at $E_0=0.8$~MeV.  Note that due to the
normalisation the (squared) core-neutron invariant mass does not present 
a simple structure directly
related to the energy of the resonance/FSI \cite{FMM01}.

The Dalitz plot for the data from the dissociation by Pb (Fig.~3, upper panel)
presents a strong $n$-$n$ FSI and a uniform density for 
$m_{\rm{nn}}^2>\sim0.5$.
Indeed, the $n$-$n$ FSI alone describes very well the projections onto both axes,
and therefore suggests that core-$n$ resonances are not present to any
significant extent. This result confirms the hypothesis of simultaneous $n$-$n$
emission employed, as described above, in the original analysis of the 
dissociation of $^{14}$Be by 
Pb \cite{FMM00}. The $r_{nn}^{RMS}$ so extracted, $5.6\pm1.0$~fm, may thus
be considered to represent
the $n$-$n$ separation in the halo of $^{14}$Be.
 
For dissociation by the C target (Fig.~3, lower panel), despite the lower
statistics, two differences are evident. Firstly, the $n$-$n$ signal is weaker,
indicating that a significant delay has occurred between
the emission of each neutron. Second, and more importantly, the agreement
between the model including only the $n$-$n$ FSI and the data for $m_{\rm{cn}}^2$
is rather poor. In order to verify whether this disagreement corresponds to the
presence of core-$n$ resonances, the core-$n$ relative energy ($E_{\rm{cn}}$) has
been explored.
It has been reconstructed for the simulations incorporating only the $n$-$n$ FSI
and compared in Fig.~4 to the data (the model calculations have been
normalized to the data above 4~MeV). For dissociation by Pb, the inclusion of
only the $n$-$n$ FSI provides a very good description of the data, with the
exception of small deviations below 1~MeV. This is in line with the Dalitz-plot 
analysis discussed above.

The deviations observed for the C target between the measured $m_{\rm{cn}}^2$
and the simulation including only the $n$-$n$ FSI clearly
correspond to structures in the $E_{\rm{cn}}$ spectrum. Moreover, these
structures are located at energies that are in line with those of states
previously reported in $^{13}$Be: the supposed $d_{5/2}$ resonance at 
2.0~MeV \cite{Ost92,Bel98,Orr00,JeanLuc} and a lower-lying 
state(s) \cite{Bel98,Orr00,Tho98,JeanLuc}.
The model-to-data ratio is about 1/2, indicating that the peaks correspond to
resonances formed by one of the neutrons in almost all decays; the solid line 
accounts for the contribution of the neutron not 
interacting with the core. If we add to the phase-space model with $n$-$n$ FSI
core-$n$ resonances ($\Gamma=0.3$~MeV) at $E_0=0.8$, 2.0 and 
3.5~MeV with intensities of 45, 35 and 20\%, respectively, the data 
are well reproduced 
(dashed line). In the case of dissociation by Pb, the lowest-lying level(s) 
appears to be present in at most 10\% of events.

In the context of the influence of the reaction mechanism, it is worthwhile 
noting that whilst some 35\% of the two-neutron removal 
cross section on the Pb target is  
attributable to nuclear 
induced breakup \cite{Lab01}, the requirement of two neutrons in coincidence 
with the
$^{12}$Be core in the present analysis reduces 
this to some 15\% -- approximately half
of the two-neutron removal cross section arises from absorption.  

By combining the information extracted from the core-$n$ channel with the $n$-$n$
correlation functions, the analysis can be extended to extract the average
lifetime of the core-$n$ resonances. If the $n$-$n$ separation in $^{14}$Be is fixed to
that obtained for dissociation by Pb, $r_{nn}^{RMS}=5.6\pm1.0$~fm, the 
delay between the emission of the neutrons $\tau_{\rm{nn}}$ needed to describe 
the $n$-$n$ correlation function for the C target may be introduced. As discussed above, 
this delay
should correspond to the lifetime of the resonances. The 
result of a $\chi^2$ analysis, represented by the dashed lines in 
Fig.~4 (bottom right panel), suggests an average lifetime of $150^{+250}_{-150}$~fm/$c$.

\section{Radiative proton capture as a probe of clustering}

In a recent 
investigation of coherent bremsstrahlung production in the
reaction $\alpha$(p,$\gamma$) it was demonstrated that the high-energy
photon spectrum is dominated by capture to form $^5$Li \cite{Hoe00}. This
result provided the motivation to 
extend the technique to probe clustering in more exotic systems \cite{Sau00}.  As 
a first test $^6$He was chosen owing to the relatively high beam intensities 
available
and the fact that structurally it is the most well established two-neutron halo
nucleus. Given a
proton wavelength of 0.7~fm at 40~MeV, 
direct capture might be observed, as a quasi-free process, on the constituents 
($\alpha$-$n$-$n$) of $^6$He
in addition to capture into $^7$Li. Moreover, the different quasi-free capture
(QFC) processes would lead to different $E_\gamma$ in the range 20--40~MeV. 

%%%%%%%%%%%%%% Setup:

Experimentally, a 40~MeV/nucleon $^6$He beam (5$\times$10$^5$~pps) was employed
to bombard a solid hydrogen target (95~mg/cm$^2$). The 
different charged reaction products  
were identified and momentum analysed using the SPEG
spectrometer. The photons were detected using 74 elements of the 
``Ch\^ateau de Cristal'' BaF$_2$ array,
with a  
total efficiency of some 70\%.  Further details including the 
analysis techniques may be found in ref. \cite{Sau00}. 
%%%%%%%%%%%%%%%%%%% 7Li+gamma

Turning to the experimental observations, the reaction
$^6$He(p,$\gamma$)$^7$Li is unambiguously identified by the $\gamma$-rays in
coincidence with $^7$Li (Fig.~5). In particular, the photon energy
spectrum, as well as the $^7$Li momentum \cite{Sau00}, is well described
assuming a $\gamma$-ray line at 42~MeV. 
The energy difference between the two particle-stable states of $^7$Li -- the 
g.s. and the first
excited state at 0.48~MeV) -- is too small for them to be
distinguished in this experiment. 
%The total efficiency for the detection of
%$^7$Li-$\gamma$ coincidences was estimated to be $37\pm2~\%$, and the deduced.
A total cross section of $\sigma=35\pm2~\mu$b was deduced. 
 
The $^6$He(p,$\gamma$)$^7$Li cross section has been calculated using a
microscopic cluster model \cite{Des95}. 
At 40~MeV, a cross section
of $\sigma=59~\mu$b was found, with 15$\mu$b going to the g.s.\ and
44$\mu$b to the first excited state \cite{Sau00}. 
The calculation was restricted to the dominant E1 multipolarity, thus leading to an
angular distribution symmetric about $90^\circ$ (Fig.~5b). The cross
section to the g.s. can also be estimated from photodisintegration measurements \cite{Sen85} 
via
detailed balance considerations and is $9.6\pm0.4~\mu$b. Given the predicted
relative populations of the ground and first excited state, a total capture
cross section of $\sigma\sim38~\mu$b is obtained, in agreement with the value
measured here.
  
%%%%%%%%%%%%%%%%%%% 6Li+gamma

QFC was investigated by searching for $\gamma$-rays in coincidence with
fragments lighter than $^7$Li. The corresponding energy spectra
(Fig.~6a,c,e) do indeed exhibit peaks below 42~MeV. In order to
establish the origin of these fragment-$\gamma$ coincidences, QFC processes on
the subsystems of $^6$He have been modelled as follows. The $^6$He projectile
was considered as a cluster ($A$) plus spectator ($a$) system in which each
component has an intrinsic momentum distribution, the corresponding energy
$E_A+E_a-m_{^6\rm{He}}$ being taken into account in the total available energy.
The reaction may be denoted as $a$+$A$(p,$\gamma$)$B$+$a$, and the $\gamma$-ray
angular distribution is assumed to be that given by the charge asymmetry of the
entrance channel \cite{Hoe99}. The intrinsic momentum distribution of
all the clusters was taken to be Gaussian in form (${\rm{FWHM}}=80$~MeV/$c$).
 In order to explore the possibility that FSI may occur in the exit channel
between the spectator, $a$, and the capture fragment, $B$, an extended version
of the QFC calculation was developed \cite{Sau00}. Here the energy in the system
$B$+$a$ is treated as an excitation in the continuum of $^7$Li, which decays in flight.
 
In the case of $^6$Li-$\gamma$ coincidences, two lines were observed
(Fig.~6a) at 30 and 3.56~MeV corresponding to the
formation of $^6$Li and the decay of the second excited state.
It was estimated that $^6$Li is
formed almost exclusively ($96^{+4}_{-24}\%$) in the 3.56~MeV excited state.
The deduced cross section was $\sigma=3.5\pm1.3~\mu$b. The lines in
Fig.~6a,b corresponds to QFC on $^5$He into
$^6$Li$^*$(3.56~MeV). The $\gamma$-ray energy spectrum is well described, whilst
the $^6$Li momentum distribution requires inclusion of $^6$Li-n
FSI. 
%The QFC calculations with fragment FSI calculations are employed in the 
%following discussions.
%%%%%%%%%%%%%%%%%%% 4He+gamma

Evidence for QFC on the $\alpha$ core, whereby the two
halo neutrons would behave as spectators,  has also been searched for. 
The photon spectrum should resemble
that observed for the $\alpha$+p reaction \cite{Hoe00}. Indeed such a
$\gamma$-ray energy spectrum (Fig.~6c) was observed.
The background, however, arising from
$^6$He breakup, in which the $\alpha$ particle is detected in SPEG and the halo
neutrons interact with the forward-angle detectors of the Ch\^ateau, is
significant. In order to minimise this background, only the backward-angle
detectors ($\theta>110^\circ$) of the Ch\^ateau were used in the analysis.
The $\gamma$-ray spectrum under this condition exhibits two components: a peak
at $E_\gamma=27$ MeV and a $1/E_\gamma$ continuum similar to coherent
$\alpha$+p bremsstrahlung \cite{Hoe00}.
 
Simulations indicate, however, that some back-scattered neutrons remain from
breakup, which would also lead to a continuous component
with a $1/E$ type spectrum in the Ch\^ateau \cite{Sau00}. 
A single background component of this form (dotted line, Fig.~6c) was 
therefore added to the QFC process
$\alpha$(p,$\gamma$)$^5$Li. The photon energy spectrum is thus well described,
as is the momentum distribution of the $\alpha$ particle. The cross section was
estimated to be $\sigma=4\pm1~\mu$b. Additional support for this interpretation 
is found in
$\alpha$-$\gamma$-n coincidences, for which 30 events were 
observed \cite{Sau00} (open symbols, Fig.~6c).

%%%%%%%%%%%%%%%%%%% d,t+gamma

Finally, d-$\gamma$ coincidences presenting a peak in the $\gamma$-ray energy
spectrum, at $E_\gamma\simeq$21~MeV, were also observed (Fig.~6e). 
The relatively low statistics arose from
the limited acceptances of the spectrometer for deuterons (Fig.~6f).
The predictions for n(p,$\gamma$)d QFC on a
halo neutron present a peak at 19~MeV (Fig.~6e) -- the small shift may
be attributable to the strong kinematic correlation between the deuteron
momentum and the photon energy, as the detection of only a small fraction of
the deuterons is
predicted \cite{Sau00}. As such no reliable estimate of the
cross section was possible.
 
There are additional QFC channels, 2n(p,$\gamma$)t and t(p,$\gamma$)$\alpha$, 
that could have been observed with finite efficiency but 
were not \cite{Sau00}. Perhaps the most interesting is QFC on the two halo
neutrons. In the case of $^6$He, several theoretical models predict the
coexistence of two configurations in the g.s. wave function: the so-called
``di-neutron'' and ``cigar'' configurations \cite{Zhu93}. Here one might expect
that the different admixtures of these could be probed by the relative strength
of the n,2n(p,$\gamma$)d,t QFC processes, whereby the corresponding free cross
sections at 40~MeV, obtained from detailed balance considerations, are
comparable: 9.6$\mu$b \cite{Ahr74} and 9.8$\mu$b \cite{Fau80}, respectively.
However, events registered in the Ch\^ateau in coincidence with tritons in SPEG
have energies below 10~MeV, whereas the 2n(p,$\gamma$)t reaction should produce
photons with $E_\gamma\approx32$~MeV. 
  
As described above, the QFC with fragment FSI model describes well the observed
monoenergetic $\gamma$-rays, as well as the momentum distribution of
the capture fragment ($B$). The $\gamma$-ray lines are associated with
specific energy distributions for the fragments in the exit channel.
Therefore, such a process will exhibit the same kinematics as capture into
continuum states above the corresponding threshold,
$^6$He(p,$\gamma$)$^7$Li$^*$$\rightarrow$$B$+$a$, provided that the equivalent
region of the continuum is populated \cite{Sau00}. If, however, all the
final states observed here were the result of radiative capture into $^7$Li,
capture via the non-resonant continuum in $^7$Li might well be expected to
occur \cite{Sid86}. This would lead to a continuous component to the
$\gamma$-ray energy spectra. Moreover, events corresponding to
$E_{^7\rm{Li}^*}=0.5$--10~MeV have not been observed in either t-$\gamma$
coincidences or $\alpha$-$\gamma$ coincidences with $E_\gamma=32$--42~MeV, nor
has the decay into $\alpha$+t for $E_{^7\rm{Li}^*}>10$~MeV. Within the picture
of QFC on clusters, this is simply explained by the absence of the
2n(p,$\gamma$)t and t(p,$\gamma$)$\alpha$ QFC processes for the $^4$He-2n
and t-t configurations, respectively.  This indicates, as suggested
by the
interferometry measurements described in the previous section,
that $^4$He-$n$-$n$ (i.e., no compact dineutron component) is the dominant 
configuration present in $^6$He$_{gs}$.

\section{Multineutron Clusters}

The very lightest nuclei have long played a fundamental r\^ole in testing nuclear
models and the underlying nucleon-nucleon interaction. 
In this
context the study of systems exhibiting very asymmetric $N/Z$ ratios may
provide new perspectives on the nucleon-nucleon interaction and few-body forces. In the
case of the light, two-neutron halo nuclei such as $^6$He, insight is already
being gained into the effects of the three-body force \cite{Zhu93}. Very
recently evidence has been presented that the ground state of $^5$H exists as a
relatively narrow, low-lying resonance \cite{Kor01}. In the case of the
lightest $N=4$ isotone, $^4$n, nothing is known despite 
experimental searches over the past 40 years \cite{Til92,Ogl89}. Theoretically
it is difficult to produce a bound 4 neutron cluster (or ``tetraneutron'') 
 \cite{Til92,Tim03,Ber03,Car02,Pie03}.  The
discovery of such neutral systems as bound states would therefore, as discussed 
by Timofeyuk \cite{Tim03} and Pieper \cite{Pie03}, have fundamental
implications for our understanding of nuclear forces. 

It is, therefore, interesting to speculate that multineutron halo nuclei
and other very neutron-rich systems may contain components 
of the wavefunction in which
the neutrons present a relatively compact cluster-like configuration.  If this
were to be the case, then the dissociation of beams of such nuclei may offer a means
to produce bound neutron clusters (if they exist) and, more generally
study multineutron correlations.  

To date the majority of searches for 
multineutron systems have relied on very low (typically $\sim$1~nb) cross 
section double-pion 
charge exchange (D$\pi$CX) and heavy-ion transfer
reactions (see, for example, refs \cite{pion,HI}).  In the case of dissociation of
an energetic beam of a very neutron-rich nucleus, relatively high cross sections
(typically $\sim$100~mb) are encountered.  Thus, even only a small component
of the wavefunction corresponding to a multineutron cluster could result 
in a measurable
yield with a moderate secondary beam intensity.  Furthermore the backgrounds
arising in D$\pi$CX and heavy-ion transfer reactions from target impurities 
and complex
many-body phase space
reactions are obviated in breakup.

The difficulty in this approach lies in the direct detection of a $^A$n cluster.
The avenue that has been explored here\footnote{Those with an historical inclination will
recognise the method as similar in principle to that employed by Chadwick \cite{Chad} 
in discovering
the neutron (thus supporting the view sometimes held by one of the authors that 
there is nothing truely new in nuclear structure physics).} is to detect 
the recoiling
proton in a liquid scintillator \cite{FMM01a}.  One of the principle advantages 
of a liquid scintallator is that neutrons may be discriminated with good efficiency from
the $\gamma$ and cosmic-ray backgrounds using pulse-shape analysis. 
Careful calibrations, employing sources, cosmic rays and the maximum 
proton recoil energy for a given $E_n$, 
permit the charge deposited and hence the energy ($E_p$) of the recoiling proton  
to be determined.
This may be compared to the energy derived from the measured time-of-flight
($E_n$):
for a single neutron and an ideal detector, $E_p$/$E_n$$\leq$1; for a 
realistic detector with 
finite resolution the limit is $\sim$1.4.  In the case of a 
multineutron cluster ($^A$n) $E_p$ can exceed the incident energy per nucleon
and
$E_p$/$E_n$ will take on a range of values 
extending beyond 1.4 --- up to $\sim$3 in the case of A=4 (Fig. 7).

The data already at hand from the study of the disociation of $^{14}$Be and 
$^{11}$Li \cite{FMM00,Lab01,FMM01}
was examined with a view to testing the method outlined above.
The details of the analyses carried out may be found in ref. \cite{FMM01a}.  The 
essential results are provided by figures 8 and 9 which display the charged
fragment particle identification (PID) derived from the Si-CsI detector 
telescope versus $E_p/E_n$.  
The $E_p/E_n$ distributions (upper panels in Figs. 8 and 9)
exhibit a general trend below 1.4: a plateau up to 1 followed by a sharp
decline, which may be fitted to an exponential distribution (dashed line). In
the region where $^A$n events may be expected to appear some 7
events with $E_p/E_n$ ranging from 1.4 to 2.2 are observed for $^{14}$Be. 
In the case
of $^{11}$Li, despite the greater number of neutrons detected (factor of 2.4),
only 4 events appear which lie just above threshold. Turning to the 
coincidences with the charged fragments, the 7 events produced by the $^{14}$Be beam fall
within a region centred on $^{10}$Be. In the case of the 4 events produced in
the reactions with $^{11}$Li no correlation appears to exist with 
any particular fragment.

The left panel in figure 10 displays in more detail the region of the
particle identification spectrum for the breakup of $^{14}$Be into lighter Be
isotopes, together with the 7 events in question. Clearly the
resolution in PID does not allow the
observed events to be unambiguously associated with a $^{10}$Be fragment. However, 
the much higher
cross-section for this channel (460$\pm$40~mb) compared to $^{11}$Be (145$\pm$20~mb)
suggests that this may be the case.
It should be noted that the PID is somewhat complicated by the fact that 
reactions also occur in the
Si-CsI telescope.  The effects of this are shown in figure 10 (right panel), whereby
the reactions in the telescope give rise to a tail extending to higher 
mass Be fragments.
A dedicated experiment including a high statistics 
target-out measurement, or ideally using a large acceptance sweeper magnet,
would eliminate this ambiguity.

As a first step towards investigating the nature of the events with $E_p/E_n>$1.4 
each was examined to verify that it corresponded to a well
defined event in both the charged
particle and neutron detectors.  Of the 7 events observed in the breakup of $^{14}$Be,
all but one survived.  
The 6 remaining events thus appear to exhibit 
characteristics consistent with detection
of a multineutron cluster from the breakup of $^{14}$Be.  Potential sources
of such events not involving the formation of a multineutron were consequently
examined \cite{FMM01a}.  
  
The most obvious source of events that may mimic a multineutron cluster is the 
detection, in the same event, of more than one neutron in the same module.
The rates at which such pile-up is expected to occur have been examined in detail
employing both simulations which reproduce the observed neutron 
angular, energy and multiplicity distributions, together with an analysis based on the
measured neutron-neutron relative angle distributions \cite{FMM01a}.
As summarised in Table~1, the two methods provide consistent results which are in 
line with
the numbers of events observed for the channels ($^{11}$Li,X+n) and 
($^{14}$Be,$^{12}$Be+n).
In the case of ($^{14}$Be,$^{10}$Be), less than one event arising from pile-up 
is estimated
to occur with $E_p/E_n>$1.4, compared to some 6 observed events.
It may be concluded, therefore, that a signal consistent with
the 
production of a multineutron cluster in the breakup of $^{14}$Be -- most 
probably in the channel $^{10}$Be+$^4$n -- has been observed at a level some 2-sigma
above that attributable to background processes. 

The average flight time of the 6 events from the target to the neutron array is
$\sim100$~ns. Unless the decay of the 4 neutron system takes place
via the emission of highly correlated neutrons, the lifetime of the 
putative tetraneutron 
is of this order or longer; suggesting a particle bound system or a
very narrow resonance.   
The conditions applied in the analysis make an estimate of the
production cross-section rather difficult. Nonetheless, if it is assumed
that these conditions affect the 
number of neutrons and $^4$n events
in a similar manner, the cross-section measured for the production
of $^{10}$Be \cite{Lab01} may be scaled by the relative yields, resulting in a
cross section 
$\sigma(\mbox{$^4$n})\sim1$~mb.

\section{Conclusions}

An experimental programme to explore clustering and correlations in
halo and related neutron-rich 
systems has been reviewed.  New approaches have been
described, including the application of neutron-neutron interferometry and Dalitz-plot
analyses to the dissociation of two-neutron halo nuclei and the investigation of 
radiative capture as a probe of clustering.  The use of these techniques to 
study very proton-rich nuclei is clearly feasible, if somewhat complicated by 
the addition of Coulomb effects. 

Present work is concentrating on a high statistics measurement of the dissociation 
of $^{6}$He \cite{E378}.  Given that $^{6}$He is structurally the 
most well known two-neutron halo system, 
this should provide a good test of the methods described here to probe correlations.   
Furthermore, correlation function analyses employing the longitudinal and transverse
neutron-neutron relative momenta \cite{soviet} should provide an independent means to 
disentangle the
halo neutron-neutron separation and time delay in emission.  Measurements completed 
very recently employing a $^{8}$He beam provided by the SPIRAL facility 
at GANIL should permit multineutron 
correlations to be explored \cite{E378}.

Turning to more exotic systems, the study of the structure of nuclei beyond the
neutron dripline has attracted renewed attention in recent years (see, for example, 
\cite{Kor01,Kor94,Kor99,Che01,Mei02}).  Beyond the core--valence neutron(s) 
correlations, which will manifest their presence as resonances or virtual states,
strong correlations may, as suggested by Seth and Parker
\cite{Set91} and Jensen and Riisager \cite{Jen91}, persist between the neutrons. 
As a first step towards exploring these ideas, high-energy single-proton removal 
reactions\footnote{Such
reactions are of considerable spectroscopic interest as the neutron configuration
of the projectile remains essantially unperturbed \cite{JeanLuc,Che01}.} with beams of
$^{6}$He \cite{Guillaume} (Fig. 11b), $^{14}$B \cite{JeanLuc} and 
$^{17}$C \cite{JeanLuc} (Fig. 11a) have been measured.  In the case of 
$t+n+n$ (Fig. 11b),
a structure similar in energy and width to that attributed previously \cite{Kor01} to 
$^{5}$H is observed.  Despite the limited statistics, a preliminary Dalitz-plot 
analysis
indicates the presence of significant $n$-$n$ correlations.  
 
As described in section 3, a number of
different reaction channels were observed in the radiative capture study. 
Beyond 
$^6$He(p,$\gamma$)$^7$Li,
evidence for quasifree capture on $^5$He, $\alpha$ and n was found. 
Of particular importance was the observation of events which
correspond to the previously measured $\alpha$(p,$\gamma$) reaction, as well as
the non-observation of capture on a dineutron. Theoretically, 
models need to be developed to describe capture on the constituent
clusters of exotic nuclei and, for comparison, capture on the projectile into
unbound final states.  In the future, the advent of intense $^{8}$He beams may
allow $\alpha$-$4n$ and $^{6}$He-$2n$ configurations to be probed.  Comparison
with the recent breakup reaction study should prove illuminating.

In terms of neutron clusters the confirmation or otherwise of the events 
observed here with a higher intensity $^{14}$Be
beam and improved fragment detection system is essential \cite{E415}.    
The search for
similar events in the breakup of $^8$He is currently underway as part of the breakup
study noted above.  Importantly, in tandem
with this experiment, a complementary study employing the d($^{8}$He,$^{6}$Li)$^{4}$n
transfer reaction, which should be sensitive to both bound and resonant states
of the $^{4}$n, was carried out \cite{Didier} (unfortunately such a 
transfer reaction study of the $^{4}$n system
with a $^{14}$Be beam is not feasible).
Searches employing other reactions such as knockout -- $^{8}$He(p,p$\alpha$)$^{4}$n 
with detection 
of the $^{4}$n and/or the proton and $\alpha$ -- are also to be encouraged.

%%%%%%%%%%%%%%%%%%%%%%%%%%%%%%%%%%%%%%%%%%%%%%%%%%%%%%%%%%%%
%%%  Acknowledgements  %%%
%%%%%%%%%%%%%%%%%%%%%%%%%%

\bigskip
\bigskip
\bigskip

We would
like to draw special attention to the key r\^oles played by
Marc Labiche ($^{14}$Be breakup) and Emmanuel Sauvan (radiative capture) 
in the work described here.
It is also a pleasure to thank the members of the E295,332,378 and 415 
collaborations and, in
particular,  
the DEMON and CHARISSA teams for their contributions. 
The support provided by the technical staffs of LPC and GANIL  
(LISE, SPEG and cyclotron 
operations crews) in preparing and executing
the experiments is gratefully acknowledged.   
This work was funded under the
auspices of the IN2P3-CNRS (France), EPSRC (United Kingdom) et FNRS (Belgique). 
Additional
support from the ALLIANCE programme (Minist\`ere des Affaires Etrang\`eres and
British Council) and the Human Capital and Mobility Programme of the European
Community (Access to Large Scale Facilities) is also acknowledged.

\newpage
  
\begin{table}
\begin{center}
 \caption{Comparison of the number of events observed (exp) with $E_p/E_n>1.4$ for each
channel with the estimated number of events expected
from pile-up. The methods are based on a Monte-Carlo simulation (sim),
and the measured relative-angle distribution of $n$-$n$ pairs (nn). 
The latter is quoted in terms of a conservative upper limit \protect\cite{FMM01a}.}
 \begin{tabular}{lllll} \noalign{\medskip}\hline\hline\noalign{\smallskip}
 Channel & $N_{\rm{2n}}^{\rm{exp}}$ & $N_{\rm{2n}}^{\rm{(sim)}}$ &
 $N_{\rm{2n}}^{\rm{(nn)}}$ \\ \noalign{\smallskip}\hline\noalign{\smallskip}
 ($^{11}$Li,X)         & 4  & $\sim$3 & $<$7.0 \\
 ($^{14}$Be,$^{12}$Be) & 0  &       0.8 & $<$1.2 \\
 ($^{14}$Be,$^{10}$Be) & 6  &       0.2 & $<$0.8 \\
 \noalign{\smallskip}\hline\hline
 \end{tabular}
\end{center}
\end{table}
 
\newpage

%%%%%%%%%%%%%%%%%%%%%%%%%%%%%%%%%%%%%%%%%%%%%%%%%%%%%%%%%%%%
%%%  Bibliography  %%%
%%%%%%%%%%%%%%%%%%%%%%%

%
\newpage

\begin{figure}
\begin{center}
\mbox{\psfig{file=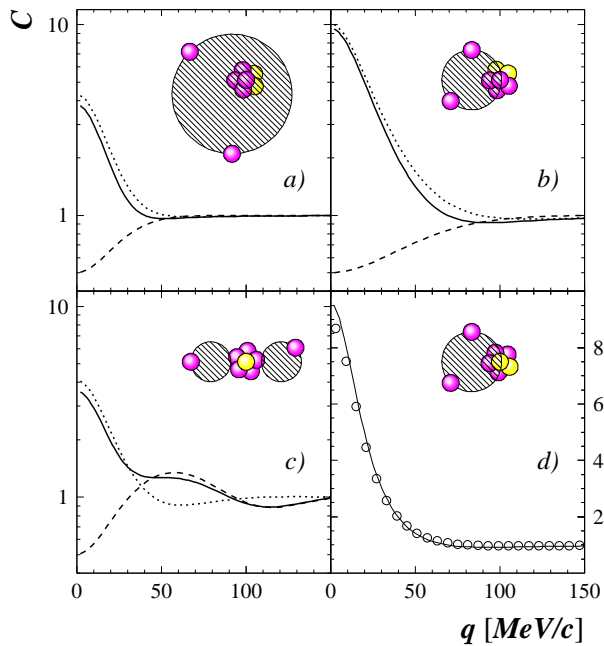,width=9cm}}
\end{center}
\caption{Neutron-neutron correlation functions, $C$, for different halo 
configurations.  The calculations are based on Gaussian sources with sizes,
$\sigma$, of (a) 6 fm, (b) 3 fm and (c) 2 fm separated by 10 fm.  The 
contributions from Fermi--Dirac statistics and the neutron--neutron FSI are 
shown by the dashed and dotted lines respectively. 
The simultaneous emission for a source size of
3 fm (solid line) is compared in (d) to a space-time extent of 3~fm, 50~fm/c
(open symbols) \protect\cite{FMM00}.}    
\end{figure}

\bigskip

\begin{figure}
\begin{center}
\mbox{\psfig{file=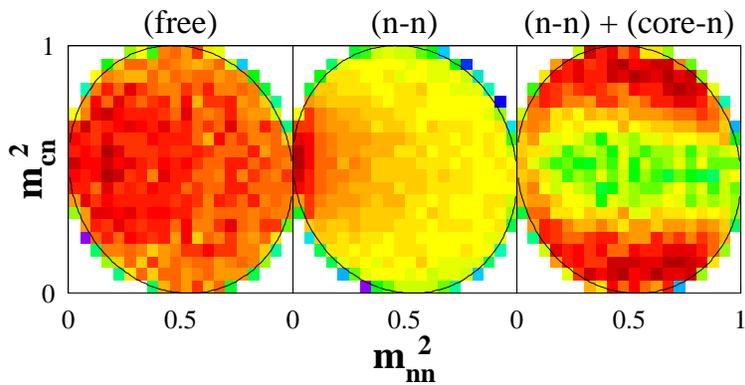,width=10cm}}
\end{center}
\caption{Dalitz plot for the simulated decay of $^{14}$Be (see text).  
In the left panel no FSI are included.  From ref. \protect\cite{NAO01}.}
\end{figure}

\newpage

\begin{figure}
\begin{center}
\mbox{\psfig{file=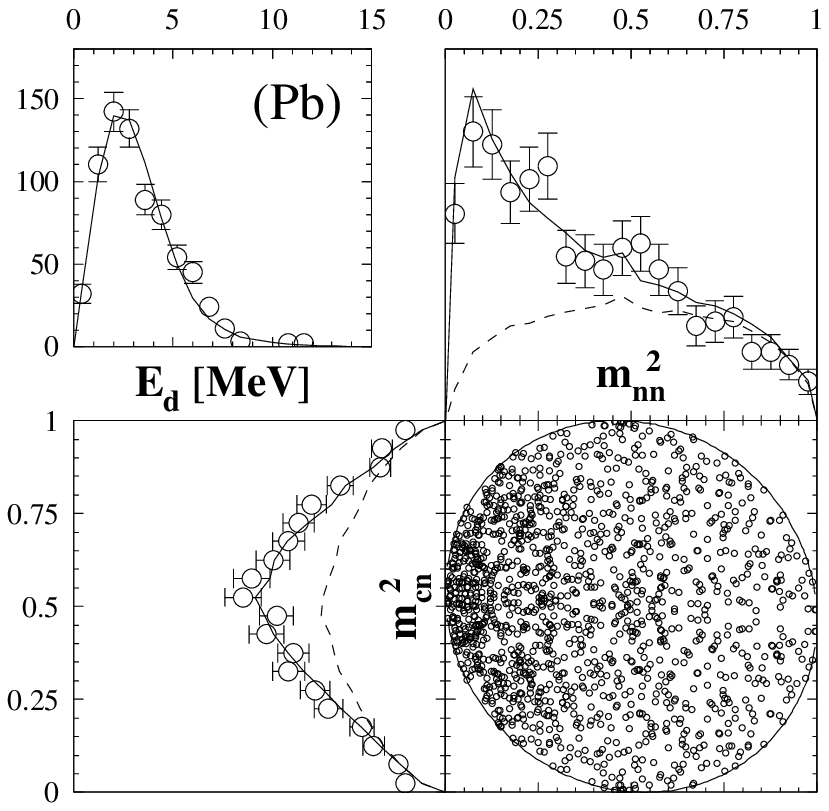,width=9cm}}
\mbox{\psfig{file=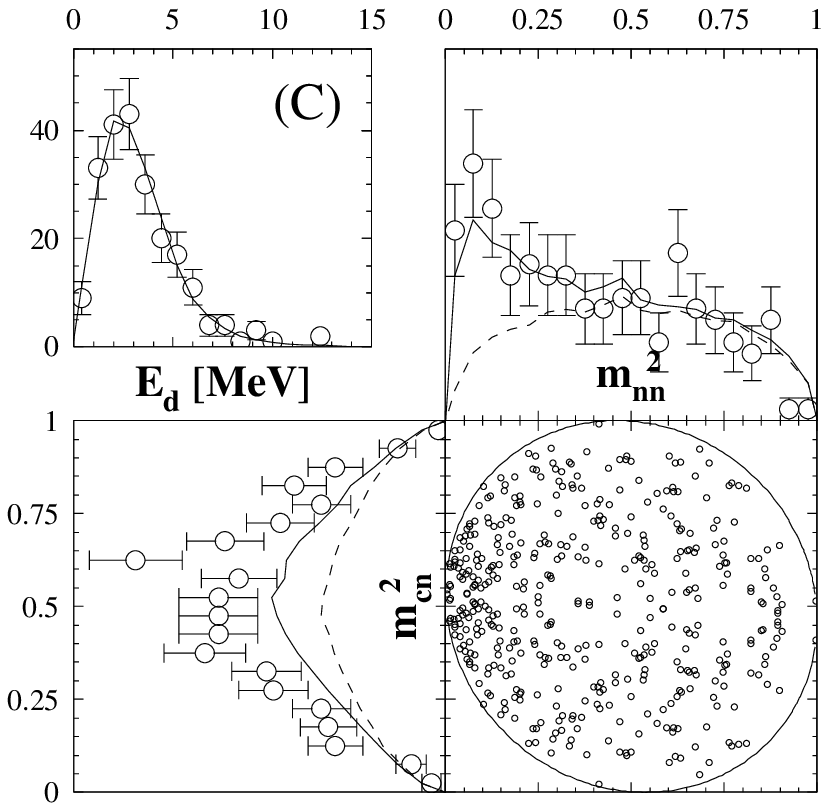,width=9cm}}
\end{center}
\caption{Dalitz plot and the projections onto the squared invariant masses
for the dissociation of $^{14}$Be by Pb (upper) and by C (lower panels). 
The lines are the 
phase-space model simulations
with/without (solid/dashed) $n$-$n$ FSI. The inset shows the measured $E_d$ spectrum.
From ref. \protect\cite{FMM01}}
\end{figure}

\newpage

\begin{figure}
\begin{center}
\mbox{\psfig{file=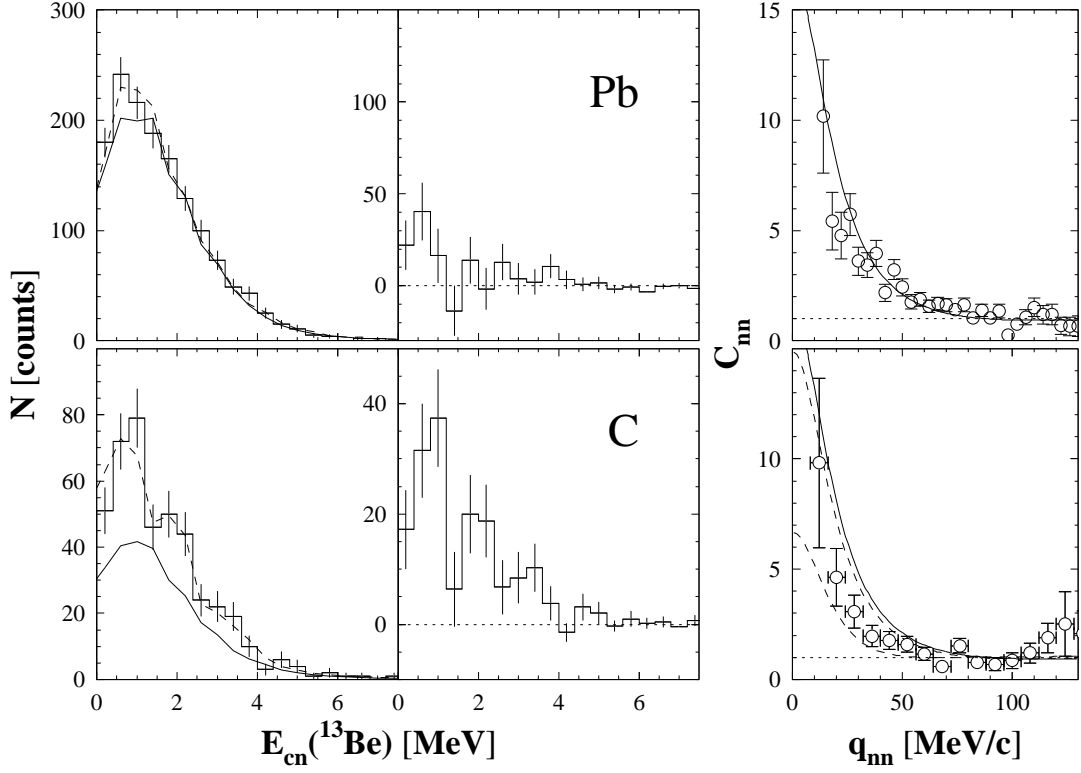,width=14.5cm}}
\end{center}
\caption{Core-$n$ relative energy distributions (left) and $n$-$n$ correlation 
functions (rightmost panels) for the dissociation of $^{14}$Be by Pb and C. The 
lines in the $E_{\rm{cn}}$ spectra are the result of the phase-space model
simulations with $n$-$n$ FSI (solid) plus core-$n$ FSI (dashed, see text). The
histograms presented in the middle panels are the difference between the data 
and the $n$-$n$ FSI simulations. The solid lines in the panels at the right are the 
$C_{\rm{nn}}$ for $r_{nn}^{RMS}=5.6$~fm and $\tau_{nn}=0$; the dashed lines 
correspond to the limits of the range $r_{nn}^{RMS}=6.6$--4.6~fm  and 
$\tau_{\rm{nn}}=0$--400~fm/$c$.  From ref. \protect\cite{FMM01}.}
\end{figure}

\newpage

\begin{figure}[tb]
 \begin{center}
  \mbox{\psfig{file=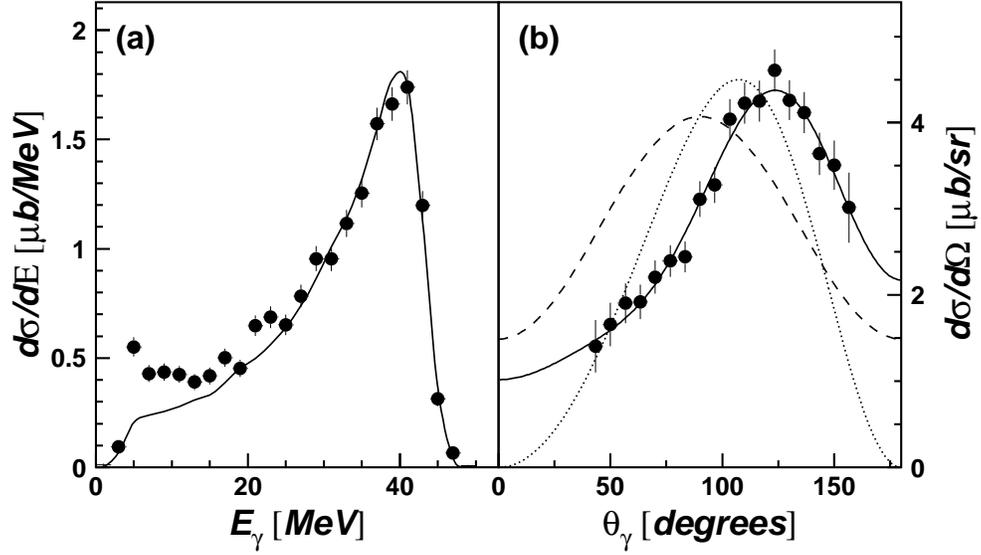,width=13cm}}
 \end{center}
 \caption{Energy (a) and angular distributions (b) in the $^6$He+p c.m.\ for
photons in coincidence with $^7$Li. The solid line in (a) is the response of
the Ch\^ateau to $E_\gamma=42$~MeV. The lines in (b) are a classical
electrodynamics calculation \protect\cite{Hoe99} (dotted), a cluster 
model \protect\cite{Sau00,Des95} (dashed),
both normalized to the data, and a Legendre polynomial 
fit \protect\cite{Wel82} (solid).
From ref. \protect\cite{Sau00}}
\end{figure}
 
\newpage

\begin{figure}
 \begin{center}
  \mbox{\psfig{file=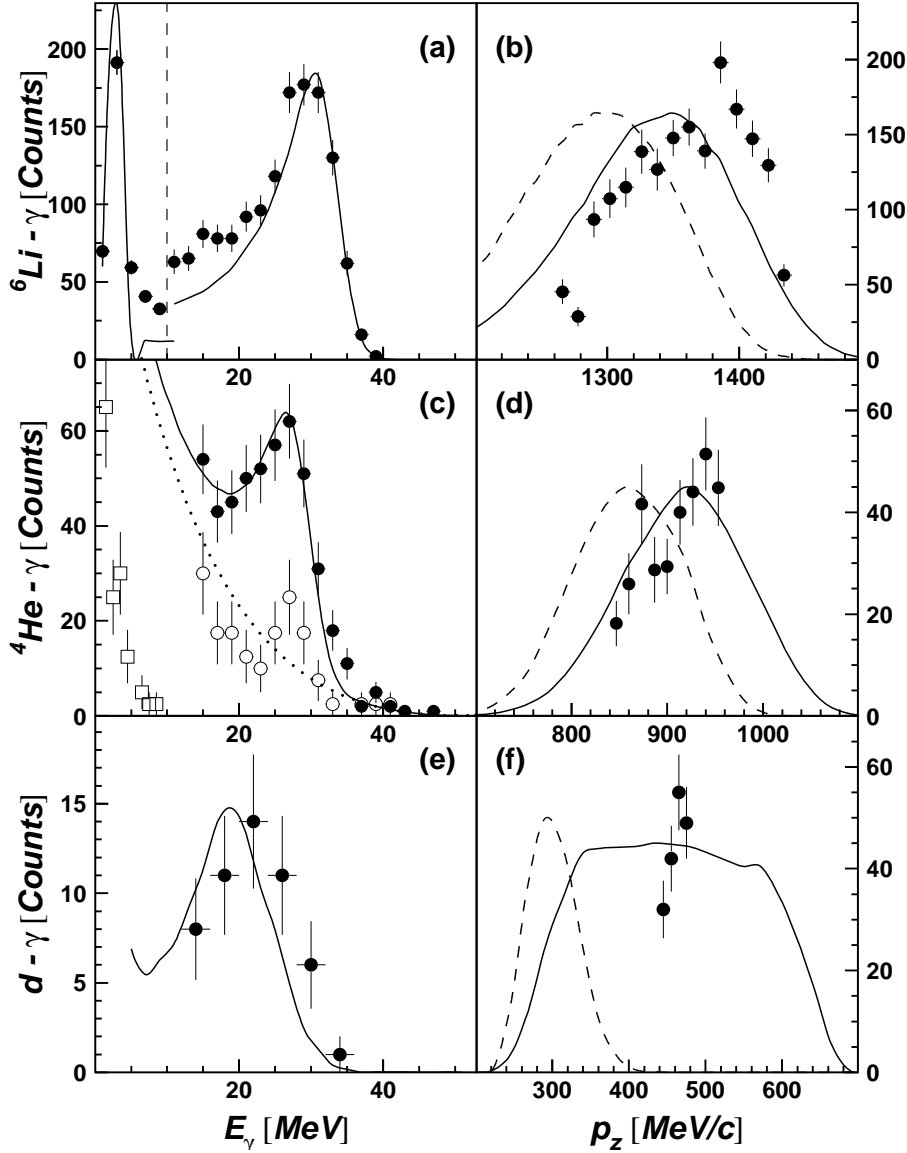,width=12cm}}
 \end{center}
 \caption{Gamma-ray energy spectrum in the $^6$He+p c.m.\ and momentum
distribution of the coincident fragment for $^6$Li (upper), $\alpha$ particles
(middle) and deuterons (lower panel). The lines correspond to calculations of
QFC on the $^5$He cluster, the $\alpha$ core and one halo neutron,
respectively; on the right with/without (solid/dashed) fragment FSI (see text).
The distribution in (a) was divided by 3 below 10~MeV, and the open symbols in
(c) are from an analysis investigating the r\^ole of the neutron background 
arising from breakup of $^{6}$He (see
ref. \protect\cite{Sau00}).}

\end{figure}

\newpage

\begin{figure}
\begin{center}
\epsfxsize=12cm
\mbox{\psfig{file=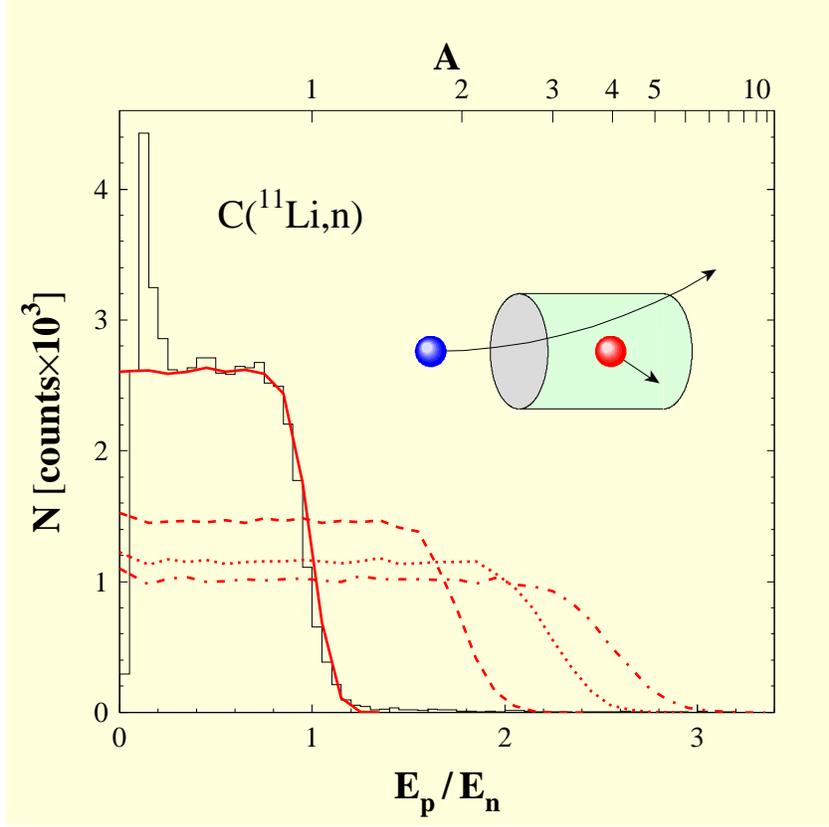}}
\end{center}
\bigskip
\caption{$E_p/E_n$ for A=1 (solid line), 2 (dashed), 3 (dotted) and 4 (dot-dashed).
In the case of A=1, comparison is made to single neutron events from the $^{11}$Li
breakup of $^{11}$Li. The excess of events at low $E_p/E_n$ arise from reactions on
the carbon component of the scintillator.  From ref. \protect\cite{NAO01}}
\end{figure}  

\newpage

\begin{figure}
\begin{center}
\mbox{\psfig{file=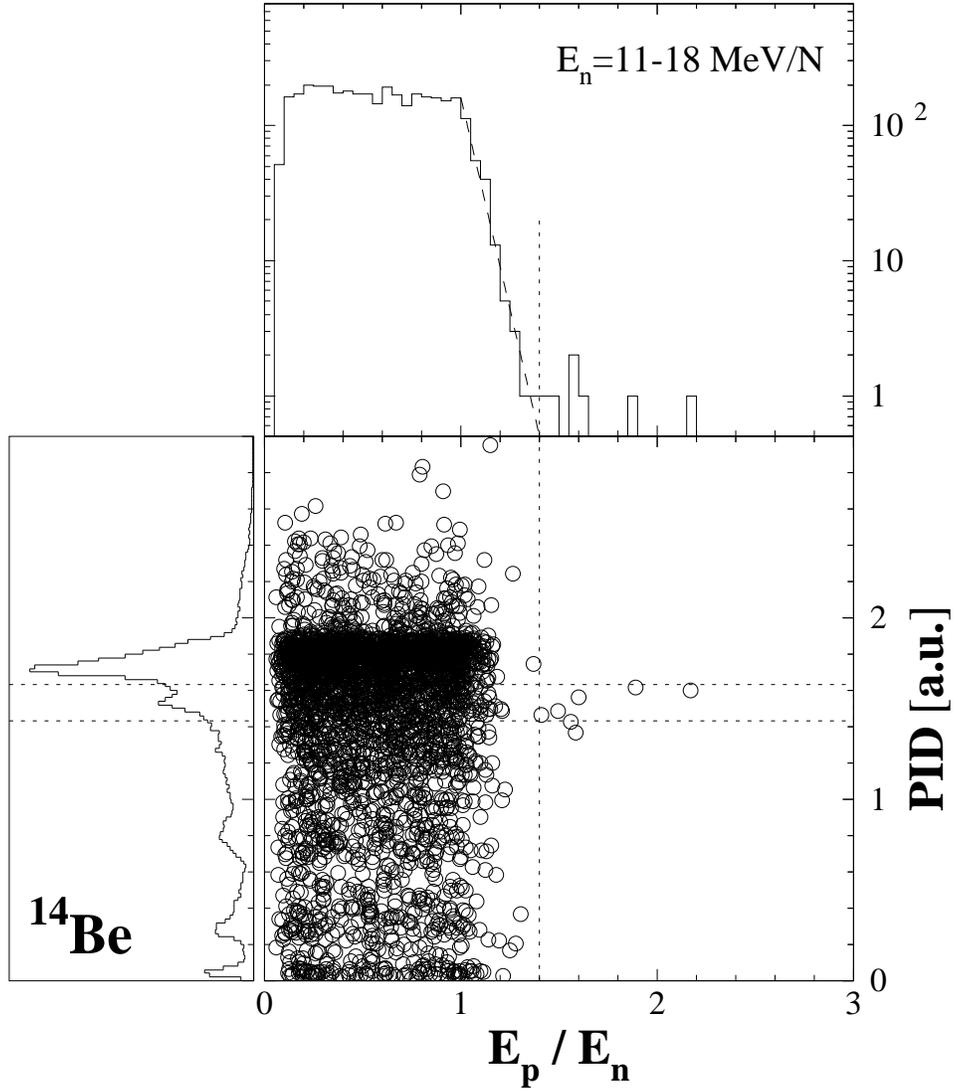,width=13cm}}
\end{center}
\caption{PID versus $E_p/E_n$ for the 
reaction ($^{14}$Be,X+n).  The prominent peak at PID$\sim$1.7 corresponds to
$^{12}$Be fragments.
The horizontal
band (dotted line) corresponds to the range of PID values encompassing the $^{10}$Be
fragments. From ref. \protect\cite{FMM01a}.}
\end{figure}

\newpage

\begin{figure}
\begin{center}
\mbox{\psfig{file=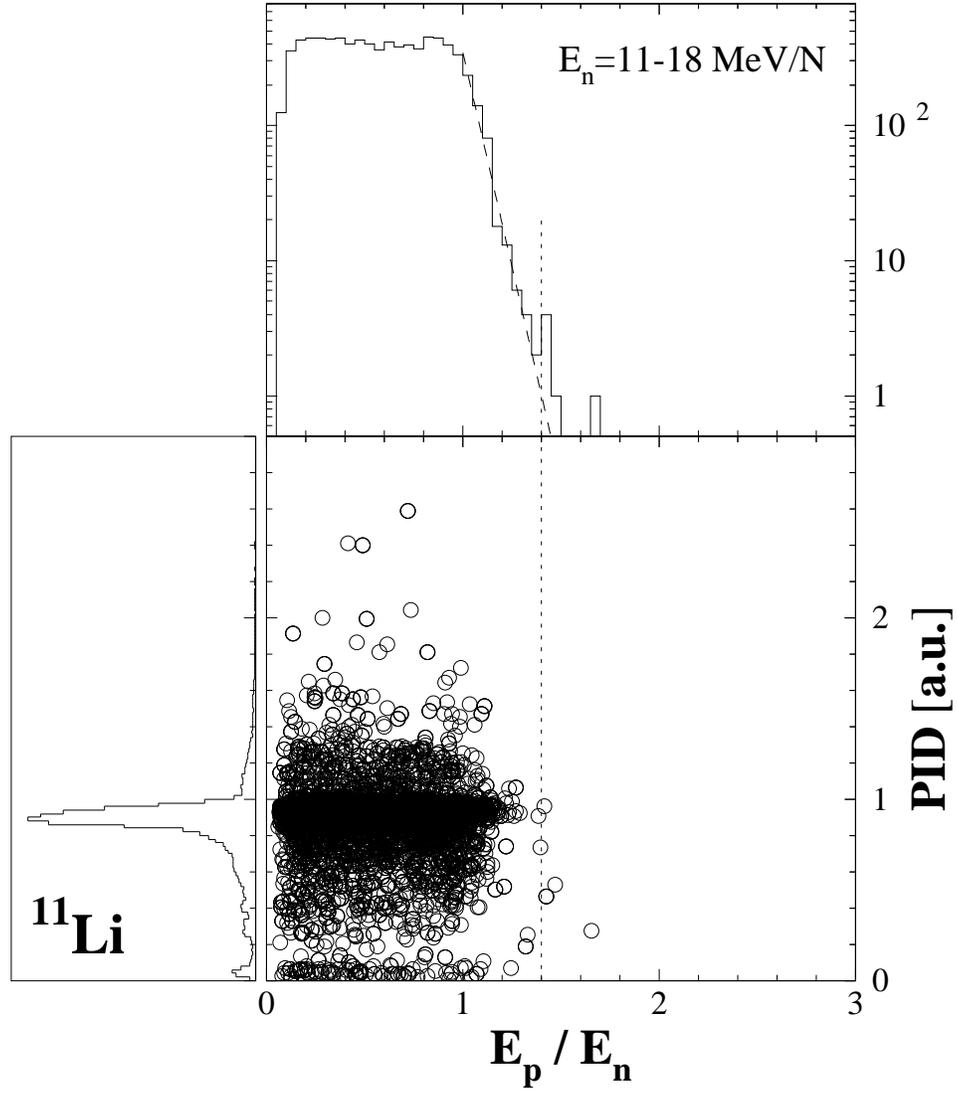,width=13cm}}
\end{center}
\caption{PID versus $E_p/E_n$ for the 
breakup of $^{11}$Li.  
The prominent peak at PID$\sim$9 corresponds to $^{9}$Li fragments 
(see, ref. \protect\cite{FMM01a}).}
\end{figure}

\newpage

\begin{figure}
\begin{center}
\mbox{\psfig{file=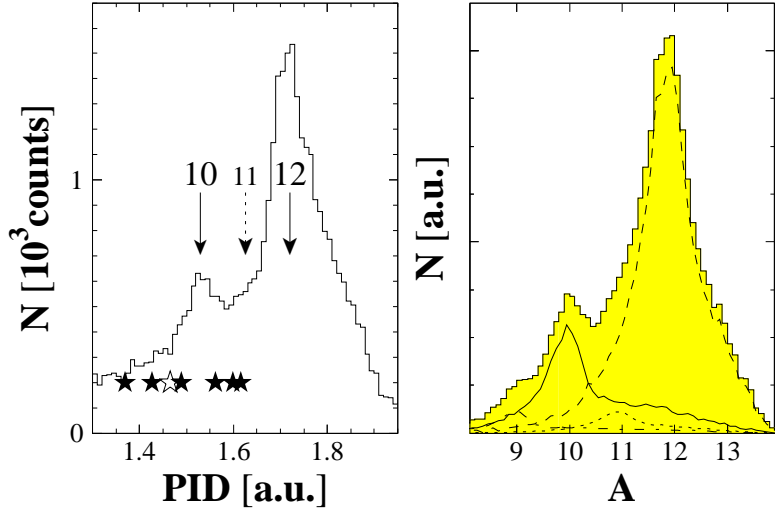,width=10.5cm}}
\end{center}
\caption{Left: detail of the particle identification spectrum around
$^{10,12}$Be for the data from the reaction ($^{14}$Be,X+n); the 7 events with
$E_p/E_n>1.4$ are denoted by the symbols. Right: results of a simulation
of the reactions ($^{14}$Be,$^{9-12}$Be) in the target and telescope; the shaded
histogram is the sum of the contributions from all four fragments
(see, ref. \protect\cite{FMM01a}).}
\end{figure}

\begin{figure}
\begin{center}
\mbox{\psfig{file=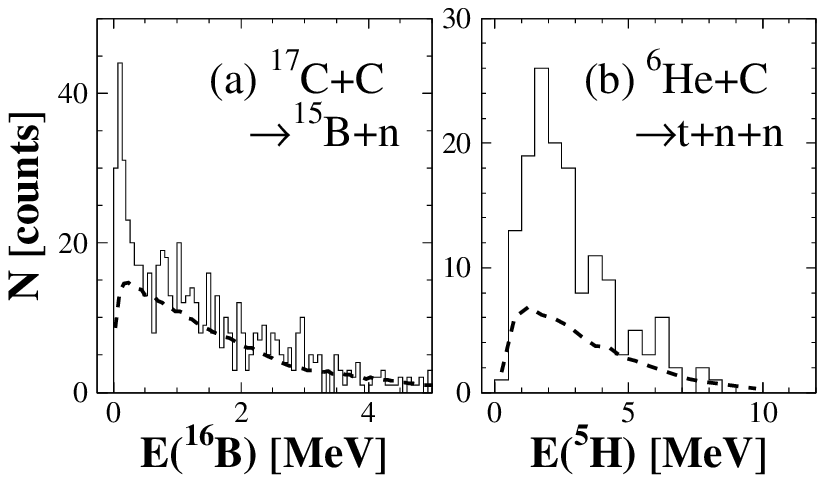,width=13cm}}
\end{center}
\caption{Decay energy spectra for (a) $^{15}B$-$n$ \protect\cite{JeanLuc} 
and (b) $t$-$n$-$n$ coincidences \protect\cite{Guillaume}
for single-proton removal reactions.  In both cases the reconstructed decay energy
spectra are compared to that expected for uncorrelated events (dashed line)  
as generated via event mixing \protect\cite{JeanLuc}.}
\end{figure}


\begin{thebibliography}{99}

\bibitem{VonO}  W. von Oertzen, H.G. Bohlen, ``Covalently Bound Molecular States in
Beryllium and Carbon Isotopes", contribution to this volume.

\bibitem{Freer}  M. Freer, ``Molecules in Nuclei'', contribution to this volume. 

\bibitem{Tanihata}  I. Tanihata, ``Halo Nuclei'', contribution to this volume.

\bibitem{Nunes} F.M. Nunes, ``Probing the Halo Structure'', contribution to this volume.

\bibitem{HBT} R.~Hanbury-Brown, R.~Twiss, Philos. Mag. {\bf 45} (1954) 663

\bibitem{Gol60} G.~Goldhaber {\em et al.}, Phys. Rev. {\bf 120} (1960) 300

\bibitem{Boa90} D.H.~Boal {\em et al.}, Rev. Mod. Phys. {\bf 62} (1990) 553

\bibitem{FMM00} F.M.~Marqu\'es {\em et al.}, Phys. Lett. {\bf B476} (2000) 219

\bibitem{E295} N.A.~Orr {\em et al.}, ``Reaction Study of the Two-Neutron Halo Nucleus
$^{14}$Be'', GANIL Proposal E295, February 1997 

\bibitem{Lab01} M.~Labiche {\em et al.}, Phys. Rev. Lett. {\bf 86} (2001) 
600; \\
M~Labiche,``Etude de la dissociation du $^{14}$Be, noyau Borrom\'een \`a halo
de deux neutrons'', Th\`ese, Universit\'e de Caen, (1999), LPCC T 99-03 

\bibitem{Mar00}F.M.~Marqu\'es {\em et al.}, Nucl. Inst. Meth. {\bf A450} (2000) 109

\bibitem{soviet} R.~Lednicky, L.~Lyuboshits, Sov. J. Nucl. Phys. {\bf 35} (1982) 770

\bibitem{NAO01} N.A.~Orr, Eur. Phys. J. {\bf A15} (2002) 109

\bibitem{FMM01}  F.M.~Marqu\'es {\em et al.}, Phys. Rev. {\bf C64} (2001) 061301R    
    
\bibitem{Dal53} R.H.~Dalitz, Philos. Mag. {\bf 44} (1953) 1068

\bibitem{Example} D.H.~Perkins, ``Introduction to High Energy Physics'', Addison Wesley
(Menlo Park, California, 1987) Chapter 4

\bibitem{Ost92} A.N.~Ostrowski {\em et al.}, Z. Phys. {\bf A343} (1992) 489

\bibitem{Bel98} A.V.~Belozyorov {\em et al.}, Nucl. Phys. {\bf A636} 419 (1998) 419 

\bibitem{Orr00} K.L.~Jones, ``The Unbound Nucleus $^{13}$Be'', Thesis, 
University of Surrey (2001); \\ 
N.A.~Orr, nucl-ex/0011002 and refs therein; \\
H. Simon, ``Aufbruchreaktionen der Halokerne $^{11}$Li und $^{14}$Be bei Relativistischen
Energien'', Thesis , Technical University of Darmstadt (1998); \\
H.~Simon {\em et al.}, in preparation

\bibitem{Tho98} M.~Thoennessen {\em et al.}, Phys. Rev. {\bf C63} 014308 (2001) 014308

\bibitem{JeanLuc} J.L.~Lecouey, ``Etudes des Syst\`emes Non Li\'es $^{13}$Be 
et $^{16}$B'',
Th\`ese, Universit\'e de Caen (2002), LPC Rapport LPCC T 02-03

\bibitem{Hoe00} M.~Hoefman {\em et al.}, Phys.\ Rev.\ Lett.\ {\bf85} (2000) 1404

\bibitem{Sau00} E.~Sauvan {\em et al.}, Phys. Rev. Lett. {\bf 87} (2001) 042501\\
		E.~Sauvan, ``Etude de la structure de noyaux riches en neutrons \`a
                l'aide de nouvelles sondes'', Th\`ese, Universit\'e de Caen, 
                LPCC T-00-01 (2000)

\bibitem{Hoe99} M.~Hoefman {\em et al.}, Nucl.\ Phys.\ A {\bf654} (1999) 779c 

\bibitem{Des95} P.~Descouvemont, Nucl.\ Phys.\ A {\bf584} (1995) 532

\bibitem{Wel82} H.R.~Weller {\em et al.}, Phys.\ Rev.\ C {\bf25} (1982) 2921

\bibitem{Sen85} M.R.~Sen\'e {\em et al.}, Nucl.\ Phys.\ A {\bf442} (1985) 215

\bibitem{Zhu93} M.V.~Zhukov {\em et al.}, Phys.\ Rep.\ {\bf231} (1993) 151

\bibitem{Ahr74} J.~Ahrens {\em et al.}, Phys.\ Lett.\ {\bf52B} (1974)  49

\bibitem{Fau80} D.D.~Faul {\em et al.}, Phys.\ Rev.\ Lett.\ {\bf44} (1980) 129

\bibitem{Sid86} S.A.~Siddiqui, N.~Dytlewski, H.H.~Thies,
                Nucl.\ Phys.\ A {\bf458}, 387 (1986)

\bibitem{Kor01} A.A.~Korsheninnikov {\em et al.}, Phys. Rev. Lett. {\bf 87} (2001) 092501

\bibitem{Til92} D.R.~Tilley, H.R.~Weller, G.M.~Hale, Nucl. Phys. {\bf A541} 
(1992) 1 and references therein

\bibitem{Ogl89} A.A.~Ogloblin, Y.E.~Penionzhkevich, in {\em Treatise on Heavy-Ion Science 
(vol.~8): Nuclei Far From Stability}, ed. D.A.~Bromley (Plenum Press, New York, 1989) 
p~261 and references therein

\bibitem{Tim03} N.~Timofeyuk, J. Phys. {\bf G29} (2003) L9; nucl-th/0301020

\bibitem{Ber03} C.~Bertulani, V.~Zelevinsky, nucl-th/0212060

\bibitem{Car02} J.~Carbonell, {\em priv. comm.}

\bibitem{Pie03} S.C.~Pieper, nucl-th/0302048 

\bibitem{pion} J.~Gr\"uter {\em et al.}, Eur. Phys. J. {\bf 4} (1999) 5

\bibitem{HI} H.G.~Bohlen {\em et al.}, Nucl. Phys. {\bf A583} (1995) 775

\bibitem{Chad}  J.~Chadwick, Nature {\bf 129} (1932) 312

\bibitem{FMM01a} F.M.~Marqu\'es {\em et al.}, Phys. Rev. {\bf C65} (2002) 044006

\bibitem{E378} F.M.~Marqu\'es, F.~Hanappe {\em et al.}, ``Multi-particle Correlations
and the Structure of Heavy He Isotopes'', GANIL Proposal E378, 
September 2000  

\bibitem{Kor94} A.A.~Korsheninnikov {\em et al.}, Phys. Lett. {\bf B326} (1994) 31;\\
A.N.~Ostrowski {\em et al.}, Phys. Lett. {\bf B338} (1994) 13

\bibitem{Kor99} A.A.~Korsheninnikov {\em et al.}, Phys. Rev. Lett. {\bf 82} (1999) 3581


\bibitem{Che01} L.~Chen {\em et al.}, Phys. Lett. {\bf B505} (2001) 21 

\bibitem{Mei02} M.~Meister {\em et al.}, Phys. Rev. Lett. {\bf 88} (2002) 102501;\\
H.G.~Bohlen {\em et al.}, Phys. Rev. {\bf C64} (2001) 024312

\bibitem{Set91} K.K.~Seth, B.~Parker, Phys. Rev. Lett. {\bf 66} (1991) 2448

\bibitem{Jen91} A.S.~Jensen, K.~Riisager, Phys. Lett. {\bf B264} (1991) 238

\bibitem{Guillaume} G.~Normand, unpublished

\bibitem{E415} F.M.~Marqu\'es, F.~Hanappe {\em et al.}, ``Search for Multineutrons and 
Correlations in the Breakup of $^{14}$Be'', GANIL Proposal E415, March 2002

\bibitem{Didier} D.~Beaumel, R.~Wolski {\em et al.}, ``Study of the 4-Neutron 
System using the d($^6$He,$^6$Li) Reaction'', GANIL Proposal E422S, March 2002

%%
\end{thebibliography}
\end{document}